\documentclass[%
preprint,
nofootinbib,
 amsmath,amssymb,
 aps,
]{revtex4}

\usepackage{cancel}
\usepackage{color}
\usepackage{slashed}
\usepackage{graphicx}% Include figure files
\usepackage{dcolumn}% Align table columns on decimal point
\usepackage{bm}% bold math
\begin{document}

\preprint{SI-HEP-2019-02}
\preprint{QFET-2019-02}

\title{Heavy neutrino searches at future $Z$-factories}% Force line breaks with \\

\author{Jian-Nan Ding$^{1}\footnote{dingjn13@lzu.edu.cn}$, Qin Qin$^{2,3}$\footnote{qin@physik.uni-siegen.de}, 
Fu-Sheng Yu$^{1,4}$\footnote{yufsh@lzu.edu.cn}}

\address{
$^1$School of Nuclear Science and Technology,  \\
Lanzhou University, Lanzhou 730000,  China\\
$^2$Theoretische Physik 1, Naturwissenschaftlich-Technische Fakult\"{a}t Universit\"{a}t Siegen, Walter-Flex-Strasse 3, D-57068 Siegen, Germany\\
$^3$School of physics, Huazhong University of Science and Technology, Wuhan 430074, China\\
$^4$Research Center for Hadron and CSR Physics, Lanzhou University and Institute of Modern Physics of CAS, Lanzhou 730000, China}

\date{\today}% It is always \today, today,

\begin{abstract}

We analyze the capacity of future $Z$-factories to search for heavy neutrinos with their mass from 10 to 85 GeV.
The heavy neutrinos $N$ are considered to be produced via the process $e^+e^-\to Z\to \nu N$ and to decay into 
an electron or muon and two jets. By means of Monte Carlo simulation of such signal events and the Standard Model 
background events, we obtain the upper bounds on the cross sections $\sigma(e^+e^-\to \nu N\to \nu\ell jj)$  
given by the $Z$-factories with integrated luminosities of 0.1, 1 and 10 ab$^{-1}$ if no signal events are observed. 
Under the assumption of a minimal extension of the Standard Model in the neutrino sector, we also present the 
corresponding constraints on the mixing parameters of the heavy neutrinos with the Standard Model leptons, 
and find they are improved by at least one order compared to current experimental constraints.

\end{abstract}

\maketitle

\section{\label{sec:level1}Introduction}

In the Standard Model (SM), only left-handed neutrinos are introduced and no mechanism is responsible for the generation 
of neutrino mass. However, the observation of neutrino oscillations \cite{Fukuda:1998mi,Ahmad:2002jz} has given the 
evidence that neutrinos have tiny but non-zero mass, which may have opened a window towards the new dynamics 
beyond the SM. To explain the origin of neutrino mass and why they are much smaller than other fermion mass, 
different kinds of seesaw mechanisms 
\cite{Minkowski:1977sc,Mohapatra:1979ia,Yanagida:1980xy,Schechter:1980gr,Magg:1980ut,Cheng:1980qt,Lazarides:1980nt,Mohapatra:1980yp,Foot:1988aq} were proposed and 
work effectively as simple and straightforward methods. 
Among them, the Type-I Seesaw \cite{Minkowski:1977sc,Mohapatra:1979ia,Yanagida:1980xy,Schechter:1980gr} 
is a quite natural extension of the SM by introducing gauge-singlet right-handed neutrinos without violating the SM 
gauge symmetries. Originally, it was proposed with the Majorana mass terms of the right-handed neutrinos at the scale 
of grand unified theories \cite{Glashow:1979nm}, which automatically lead to tiny neutrino mass. 
Later, it was found that a much lower Majorana mass scale, \textit{e.g.} $\mathcal{O}$(10) GeV, is also possible to explain 
the neutrino mass (see \textit{e.g.} \cite{Asaka:2005pn,Asaka:2005an}), given small Dirac mass terms or some 
symmetry-protected cancellations in the neutrino mass matrix 
\cite{Appelquist:2002me,Appelquist:2003hn,Shaposhnikov:2006nn,Kersten:2007vk,Moffat:2017feq}. 
On the way to verify any of the Type-I Seesaw models, the most important evidence would be direct discoveries of 
heavy neutrinos. In this sense, the low-scale seesaw models are of special interest, because their particle spectra 
may contain heavy neutrinos with mass at $\mathcal{O}$(10) GeV, within the reach of the 
colliders running now or in the near further. For example, extending the SM by three right-handed neutrinos 
with mass smaller than the electroweak scale \cite{Asaka:2005pn,Asaka:2005an}, three heavy neutrinos beyond 
the SM spectrum are generated, one of which has a mass at keV scale as a dark-matter candidate 
and the other two at GeV to hundred GeV scale. 
Another interesting property of this model is that the mixing between the light neutrinos and the heavy neutrinos 
can be quite large. Therefore, heavy neutrinos in such a model have good opportunities to be detected by collider
experiments. Once they are detected, it will also give us hints about leptonic CP violation as discussed in \cite{Caputo:2016ojx}.

Experimentally, there have been direct searches for heavy neutrinos with $\mathcal{O}$(10) GeV mass by the DELPHI 
Collaboration \cite{Abreu:1996pa}. In their searches, the heavy neutrinos were considered to be produced 
via the $e^{+}e^{-}\to Z \to \nu N$ process and decay into visible final states. Unfortunately, no signals were observed 
and thus only upper bounds on mixing parameters were given. There has also been a direct search 
by the CMS collaboration \cite{Sirunyan:2018mtv}. On the other hand, such heavy neutrinos also 
receive constraints from indirect searches like neutrinoless double-beta ($0\nu2\beta$) decays 
\cite{Elliott:2004hr,Benes:2005hn,Rodejohann:2011mu}, and can be explored in meson decays and $\tau$ 
decays \cite{Atre:2009rg}. Regarding future, there are several lepton colliders proposed by different 
communities, including the Circular Electron Positron Collider (CEPC) \cite{CEPCStudyGroup:2018ghi}, the 
International Linear Collider (ILC) \cite{Baer:2013cma}, the FCC-ee (formerly known as the TLEP) 
\cite{Gomez-Ceballos:2013zzn} and the Super $Z$ Factory. While the main target of most of these colliders is precision 
study of the Higgs boson properties, they will also be capable of searching for some new particles, new dynamics 
and even of studies of quantum chromodynamics and hadrons (see \textit{e.g.} 
\cite{Qin:2017aju,Dev:2017ftk,Li:2018cod,Ali:2018ifm,Cheng:2018khi}). Most importantly here, they will be ideal 
facilities to search for heavy neutrinos. Such abilities of the CEPC with $\sqrt{s}$ = 240 - 250 GeV and of the ILC 
with $\sqrt{s}$ = 1 TeV have been investigated by \cite{Liao:2017jiz,Zhang:2018rtr}. One can refer to e.g. 
\cite{delAguila:2008cj,Das:2012ze,Degrande:2016aje,Das:2016hof,Abada:2018sfh,Das:2018usr,Pascoli:2018heg} 
for collider searches for neutrinos with higher masses.

In this work, we will focus on searching for heavy neutrinos $N$ with $\mathcal{O}$(10) GeV mass at future 
$Z$-factories with integrated luminosities of 0.1, 1 and 10 ab$^{-1}$. Similar to DELPHI, we will also consider 
the signal production process $e^+e^-\to Z\to \nu N$ of heavy neutrinos, and reconstruct the heavy neutrinos 
by their visible decaying final states each containing one charged lepton and two quark jets. The background mainly
originates from the SM processes $e^+e^-\to jjjj$, $\tau^+\tau^-$ and $b\bar{b}$. Compared to a Higgs factory such as the 
CEPC with $\sqrt{s}$ = 240 - 250 GeV, we find that future $Z$-factories are much more sensitive to heavy 
neutrinos with mass below 80 GeV, because the $N$ production cross sections at the $Z$-mass pole is typically 
higher than those at 240 - 250 GeV by orders. We have noticed that employing the technique of displaced-vertex
detection \cite{Antusch:2016vyf,Antusch:2017pkq}, a search for heavy neutrinos at a $Z$-factory is almost free 
of background and can thus set even more stringent constraints to the mixing parameters between the heavy 
neutrinos and the SM leptons than a normal search. However, the prerequisite of the displaced-vertex technique depends on a strong 
assumption that the studied heavy neutrinos have lifetimes long enough to fly a detectable distance before decaying, 
which is not true for many models. For example, if a heavy neutrino has decay channels with large decay widths, 
then it will have a very short lifetime so that its decay vertex is "not displaced". Heavy neutrinos in a model with 
a Majoron $J$ \cite{Chikashige:1980ui,Schechter:1981cv}, which was introduced to generate the Majorana scale 
in an ultraviolet-complete way, have such a feature. Typically, they can efficiently decay into the light Majoron via the 
invisible channels $N\to\nu J$ and thus have short lifetimes. Therefore, while the constraints given by 
\cite{Antusch:2016vyf,Antusch:2017pkq} do not apply to such heavy neutrinos, those given in this work are still valid. 
In other words, the results of our work without making use of the displaced-vertex technique are more model 
independent than \cite{Antusch:2016vyf,Antusch:2017pkq}.

The rest of the paper is orgnized as follows. In Section \ref{sec:level2}, we will introduce the general setup of the 
new-physics scenario that we consider. The simulation of the production and decay processes of the heavy neutrinos 
and the corresponding background events at $Z$-factories will be described, and the event selection conditions will 
be explained in Section \ref{sec:level3}. 
We will present the results in Section \ref{sec:level4} and conclude by Section \ref{sec:level5}.

%In this paper, we study the signal and background of right-handed heavy neutrino with mass about dozens of GeV at CEPC \cite{CEPCStudyGroup:2018ghi} and super Z factory \cite{15}. Next section we will briefly discuss the low energy scale heavy neutrino and its mixing parameters. Then we want to discuss the phenomenological analysis at future electron positron colliders. For experiment, we hunt for the heavy neutrino with dozens of GeV mass and which will mix with three generation standard model neutrino. We will study the model independent cross section and constrain the mixing parameters in exactly model. We give a conclusion at the end of article.

\section{\label{sec:level2}General setup of the scenario}

In this work, we consider a scenario generating the tiny neutrino mass via the Type-I seesaw mechanism but with 
a low Majorana mass scale (see \textit{e.g.} \cite{Asaka:2005pn,Asaka:2005an}). It introduces $n$ 
right-handed neutrinos $R_{j}$ ($j$ = 1, ..., $n$), which are singlets of the SM gauge group SU(2), 
and their kinetic terms, mass terms and interaction terms with the SM fields are given by 
\begin{align}\label{eq:lagrangian}
\mathcal{L} \ni {1\over2}\sum_j\overline{R}_{j}i\slashed{\partial}R_{j}-\sum_{i,j}y_{ij}\overline{L}_{i}\widetilde{H} R_{j} 
- \frac{1}{2}\sum_j\overline{R}_{j}^{c}M_{R}R_{j}+h.c.\ ,
\end{align}
where $\widetilde{H}=i\tau_{2}H^{*}$, the lepton SU(2)$_L$ doublet $L = (\nu_{\ell L},\ell_L)^\text{T}$ with $\ell = e,\mu,\tau$, 
$y$ is the $3\times n$ Yukawa coupling matrix and the $n\times n$ Majorana mass matrix $M_R$ is generated by 
some high-scale dynamics. After the spontaneous symmetry breaking of the Higgs field, we can diagonalise 
the neutrino mass matrix and obtain the $3+n$ mass-eigenstate neutrinos, including 3 light neutrinos 
$\nu_i$ and $n$ heavy neutrinos $N_j$\footnote{For clarification, here $N_j$ is not a Majorana neutrino by itself, but the left-handed 
component of a Majorana neutrino. It corresponds to $N_j^c$ in \cite{Atre:2009rg}.}. Now, a flavor eigenstate is a superposition 
of the mass eigenstates, as 
\begin{align}
\nu_{\ell} = \sum_{i=1}^{3}U_{\ell i}\nu_{i} + \sum_{j=4}^{3+n}V_{\ell j}N_{j},\qquad \text{with}\ \ell = e, \mu, \tau, 
\end{align}
and thus the neutrino-relevant weak interaction terms are given by 
\begin{align}\label{eq:mixing}
\mathcal{L} \ni & -\frac{g}{2\cos\theta_{W}}Z_{\mu}\sum_{\ell} \left(\sum_{i=1}^{3}U^{*}_{\ell i}\overline{\nu}_{i} + \sum_{j=4}^{3+n}V^{*}_{\ell j}\overline{N}_{j}\right) \gamma^{\mu}P_{L} \left(\sum_{i'=1}^{3}U_{\ell i'}{\nu}_{i'} + \sum_{j'=4}^{3+n}V_{\ell j'}{N}_{j'} \right)\nonumber\\
&-\left[\frac{g}{\sqrt{2}}W^{+}_{\mu}\sum_{\ell}\left(\sum_{i=1}^{3}U^{*}_{\ell i}\overline{\nu}_{i}\gamma^{\mu}P_{L}\ell 
+\sum_{j=4}^{3+n}V^{*}_{\ell j}\overline{N}_{j}\gamma^{\mu}P_{L}\ell\right) +h.c. \right] \, .
\end{align}
It has been proved by \cite{Lee:1977tib} that the non-diagonal weak currents here do not vanish because $\nu_\ell$ 
and $R_j$ are different representations of the weak SU(2) group.

\begin{figure}[!ht]
\includegraphics[width=0.5\textwidth]{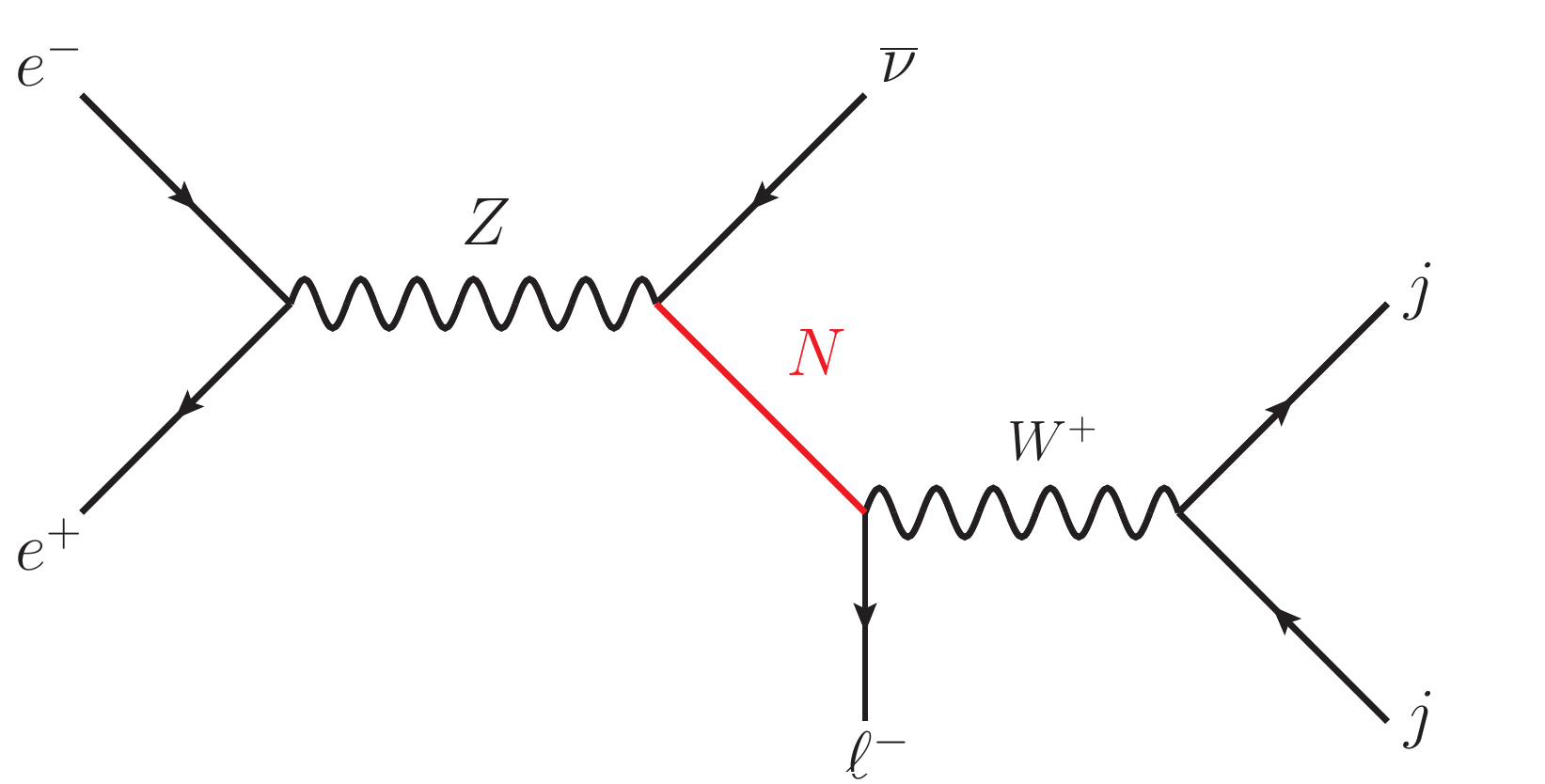}
\caption{The Feynman diagram of the signal with $\ell=e$ or $\mu$. }\label{fig:signal}
\end{figure}

From the above interaction terms \eqref{eq:mixing}, we read out that a heavy neutrino (with the mass 
smaller than the $Z$ boson mass) can be produced associated with light neutrinos via $e^+e^-\to Z \to \overline{\nu}N(\nu\overline{N})$, 
which is actually the dominant production process.\footnote{Even when $Z\to N\overline{N}$ is kinematically allowed, 
the cross section is doubly suppressed by the tiny mixing matrix elements $|V_{\ell j}|^2$, which receive stringent constraints 
from previous experiments such as the DELPHI \cite{Abreu:1996pa}.} We also find that the produced heavy neutrino 
can decay weakly into one charged lepton and one on-shell or off-shell $W$ boson up to its mass, $N\to \ell^-W^{+(*)}$. 
Here, to reconstruct the heavy neutrinos, we choose the signal events with the decaying products including one charged 
lepton and two jets, all of which can be collected by detectors, as displayed in FIG. \ref{fig:signal}. The kinematics 
requires that the mass of the heavy neutrinos are smaller than the $Z$ boson mass.

\begin{figure}[!ht]\label{fig:xsection}
\includegraphics[width=0.51\textwidth]{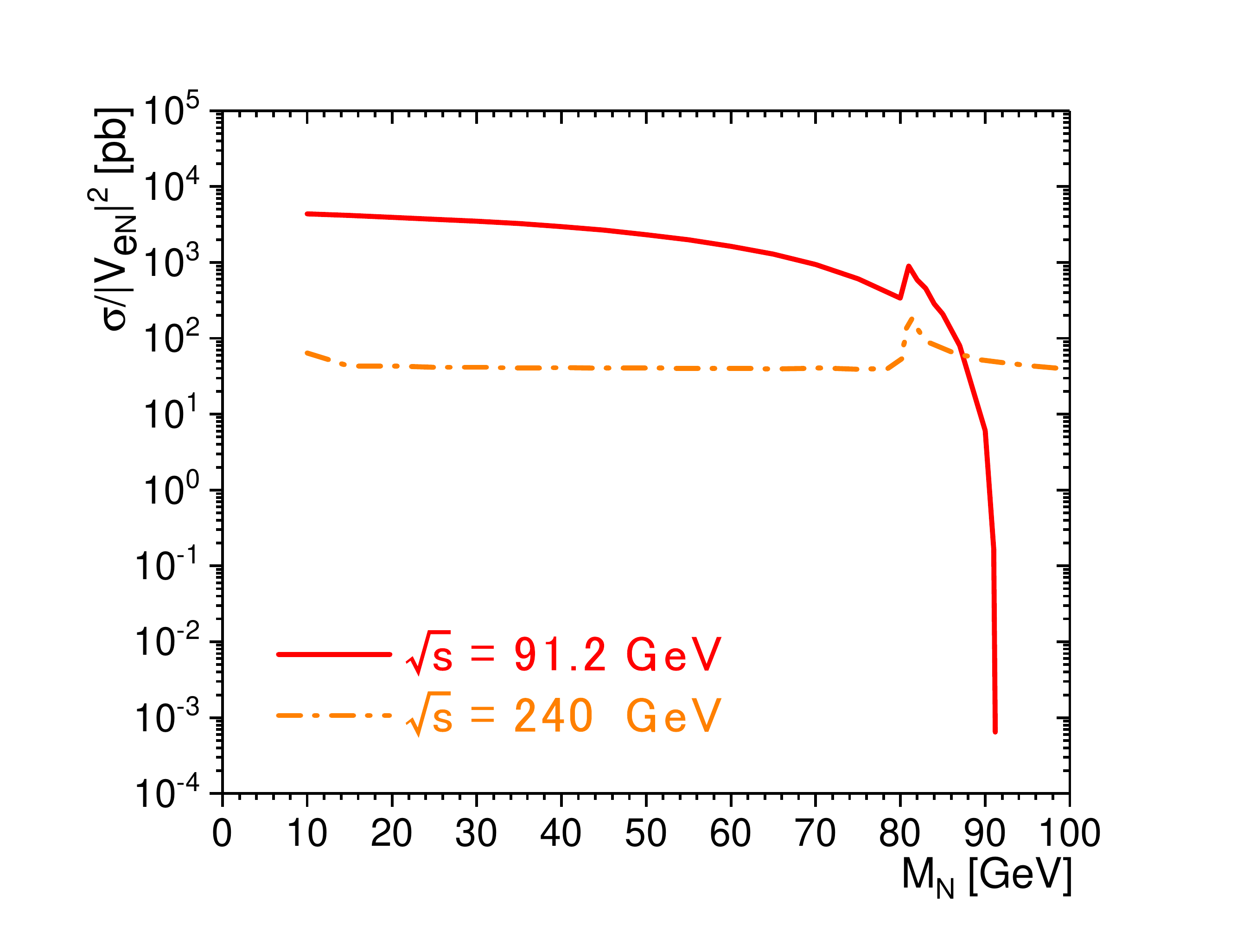}\hspace{-0.5cm}\includegraphics[width=0.51\textwidth]{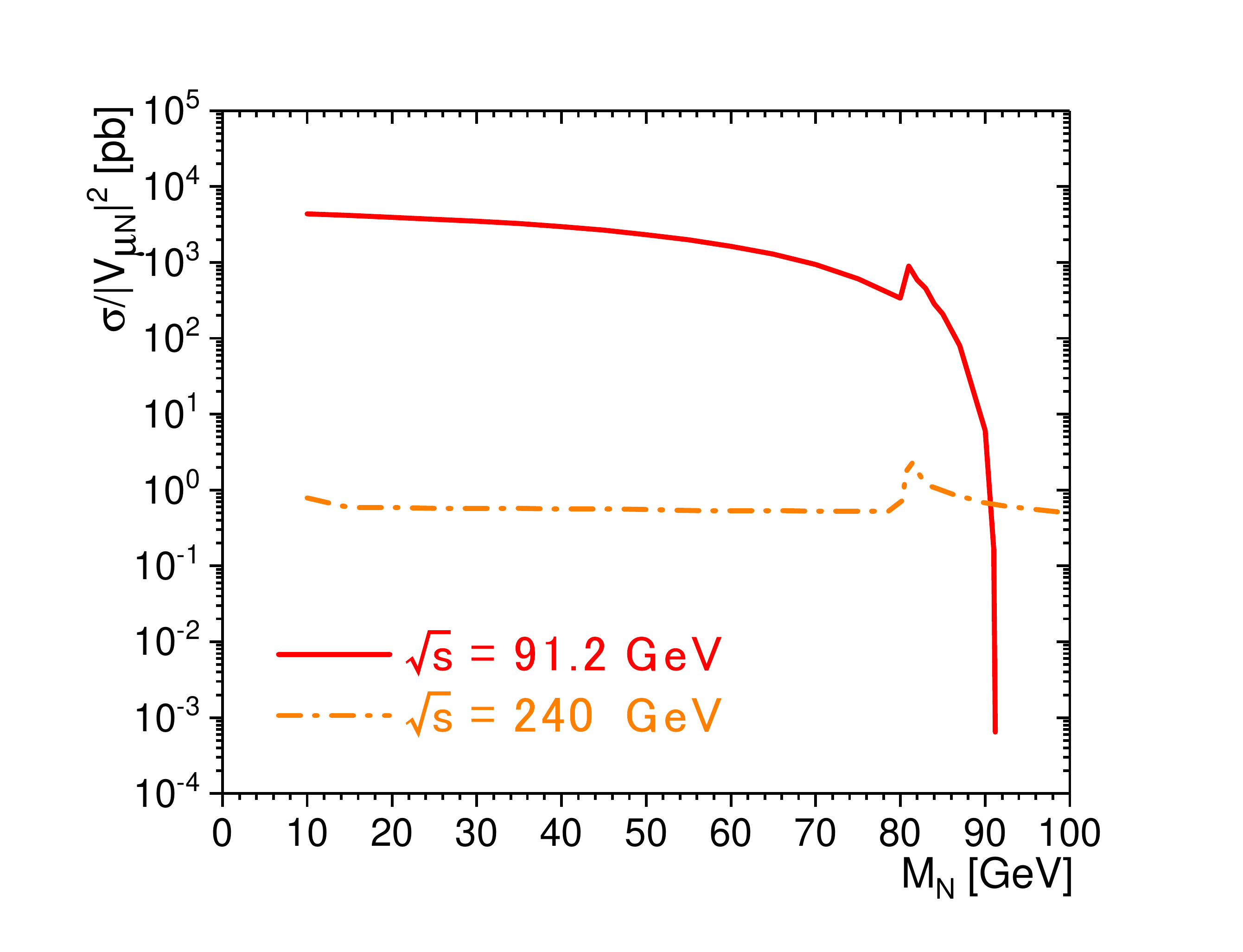}
\vspace{-0.5cm}
\caption{The production cross sections of $e^{+}e^{-}\to\nu N \to \nu e jj$ (left) and 
$e^{+}e^{-}\to\nu N \to \nu\mu jj$ (right) at a $Z$-factory (red) and a Higgs factory (orange) 
with the heavy neutrino mass $M_N$ varying from 10 to 100 GeV.}
\end{figure}

To get a hint of the potential of a $Z$-factory in searching for such heavy neutrino signals, we compare the 
performances of a $Z$-factory and a Higgs factory with the electron-positron 
collision energy at $M_Z=$ 91.2 GeV and 240 GeV, respectively. Under the assumption of the new dynamics coming only 
from the terms in \eqref{eq:lagrangian}, the dependence of the signal production cross sections 
$\sigma({e^{+}e^{-}\to \nu\ell jj})$\footnote{By $\nu\ell$ we always sum over all possible leptons and their antiparticles, 
including both the lepton-number-conserving final states like $\bar{\nu}\ell^-$ and the lepton-number-breaking 
ones like $\nu\ell^-$.}
on the corresponding mixing parameters $|V_{\ell N}|^{2}$ is obtained and shown in FIG. \ref{fig:xsection}, with 
$\ell = e$ and $\mu$. In either the $\ell=e$ or $\ell=\mu$ case, we consider only the heavy neutrino mixing 
with the electron or the muon sector, such that the  production cross sections only depend on the mixing 
parameters $|V_{e N}|^{2}$ or $|V_{\mu N}|^{2}$. The results are calculated by MadGraph \cite{Alwall:2011uj} 
with the new dynamics implemented via FeynRules \cite{Alloul:2013bka,Christensen:2008py,Degrande:2011ua}. We find that 
if the heavy neutrino mass 10 GeV $< M_N < $ 80 GeV, the signal cross sections at a $Z$-factory exceed  
$10^3 |V_{\ell N}|^{2}$ fb, while the cross sections at a Higgs factory are smaller by 1 to 2 orders in 
the $\ell = e$ case and by 3 to 4 orders in the $\ell = \mu$ case. 
The searching for heavy neutrinos with the mass between 10 GeV and 85 GeV at a $Z$-factory is obviously 
better than that at a Higgs factory with the center-of-mass energy of 240 GeV and a similar luminosity.

Next, we emphasize that even we switch on the mixing between heavy neutrinos and all the other lepton 
sectors, the signal cross sections $\sigma({e^{+}e^{-}\to \nu N\to \nu\ell jj})$ at a $Z$-factory still basically only depend on 
the one corresponding mixing parameter $|V_{\ell N}|^2$. The reason is as follows. 
The cross section $\sigma({e^{+}e^{-}\to \nu N_k})$ summing up three light neutrinos and their anti-particles is proportional to 
the mixing parameter $\sum_{i=1}^3|(U^\dagger V)_{ik}|^2$, and under the limit that the matrix $U$ is almost unitary, we have 
$\sum_{i=1}^3|(U^\dagger V)_{ik}|^2 \approx \sum_{\ell}|V_{\ell k}|^2$. On the other hand, safely neglecting the difference between 
the charged lepton mass in $N_k$ decays, we have that $\mathcal{B}(N_k\to\ell jj)\propto |V_{\ell k}|^{2}/\sum_{\ell'}|V_{\ell'k}|^{2}$. 
Therefore, the two $\sum_{\ell}|V_{\ell j}|^{2}$ factors get cancelled when we multiply $\sigma({e^{+}e^{-}\to \nu N_k})$ 
by $\mathcal{B}(N_k\to\ell jj)$, and the relation $\sigma({e^{+}e^{-}\to\nu N_k\to \nu \ell jj})\propto|V_{\ell k}|^{2}$ is obtained. 
Here, we emphasize that unless specially noted, our analysis in this work will not depend on the assumption of this 
paragraph and the previous paragraph, which means that the validity of the results of this work is not limited to the specific model extending 
the SM with only the \eqref{eq:lagrangian} terms.

The small peaks appearing where $M_N$ is slightly above the $W$ 
boson mass are due to the opening of $N$ decaying into an on-shell $W$ boson. The decline of the cross sections  at the mass close to 90 GeV results from the suppression of the phase space.

%\cite{nuMSM1,nuMSM2}

%\begin{align}\ -\mathcal{L}^{\nu}_{mass} = \frac{1}{2}\left( \sum_{m=1}^{3}m_{D}\overline{\nu}_{mL}\nu_{mR}^{c} + \sum_{m'=4}^{3+n}M_{N}\overline{N}_{m'L}^{c}N_{m'R} \right)\end{align}

%Actually, the neutrino oscillation experiment require that it have to include 2 right-handed neutrino at least. On the other hand, if the number of heavy neutrino is more than 3, it will include some free parameters which can not be constrained by measurement. Usually, peopler consider the case with 2 or 3 heavy neutrinos, which depend on whether there is a massless neutrino in nature.  Here, we consider the heavy neutrinos case with the number to be 2.

%And then, we can write the mass eigenvalue of light neutrino mass:\begin{align}\  m_{\nu}={m_{D}}M_{N}^{-1}m_{D}^{T}\end{align}where $m_{\nu}$ is $2\times2$ diagonal matrix. And for lighter mass of neutrino, the scale of heavy neutrino can be dozens of GeV.

\section{\label{sec:level3}Event simulation and selection}

For event simulation, we use MadGraph \cite{Alwall:2011uj} as the event generator for both the SM background and the 
signal with the new dynamics implemented via FeynRules \cite{Alloul:2013bka,Christensen:2008py,Degrande:2011ua} as mentioned previously. 
After that, Pythia8 \cite{Sjostrand:2014zea} and Delphes \cite{deFavereau:2013fsa} are used for further hadronization and fast detector simulation, 
respectively. In the detector simulation, the $eekt$ algorithm and $exclusive$ search have been chosen to construct jets, which, 
compared to the default $antikt$ algorithm, is more efficient for a lepton collider and mitigates energy peak drifts of jets. 

\begin{figure}[h]
\begin{minipage}[c]{0.5\textwidth}
\includegraphics[width=1.0\textwidth]{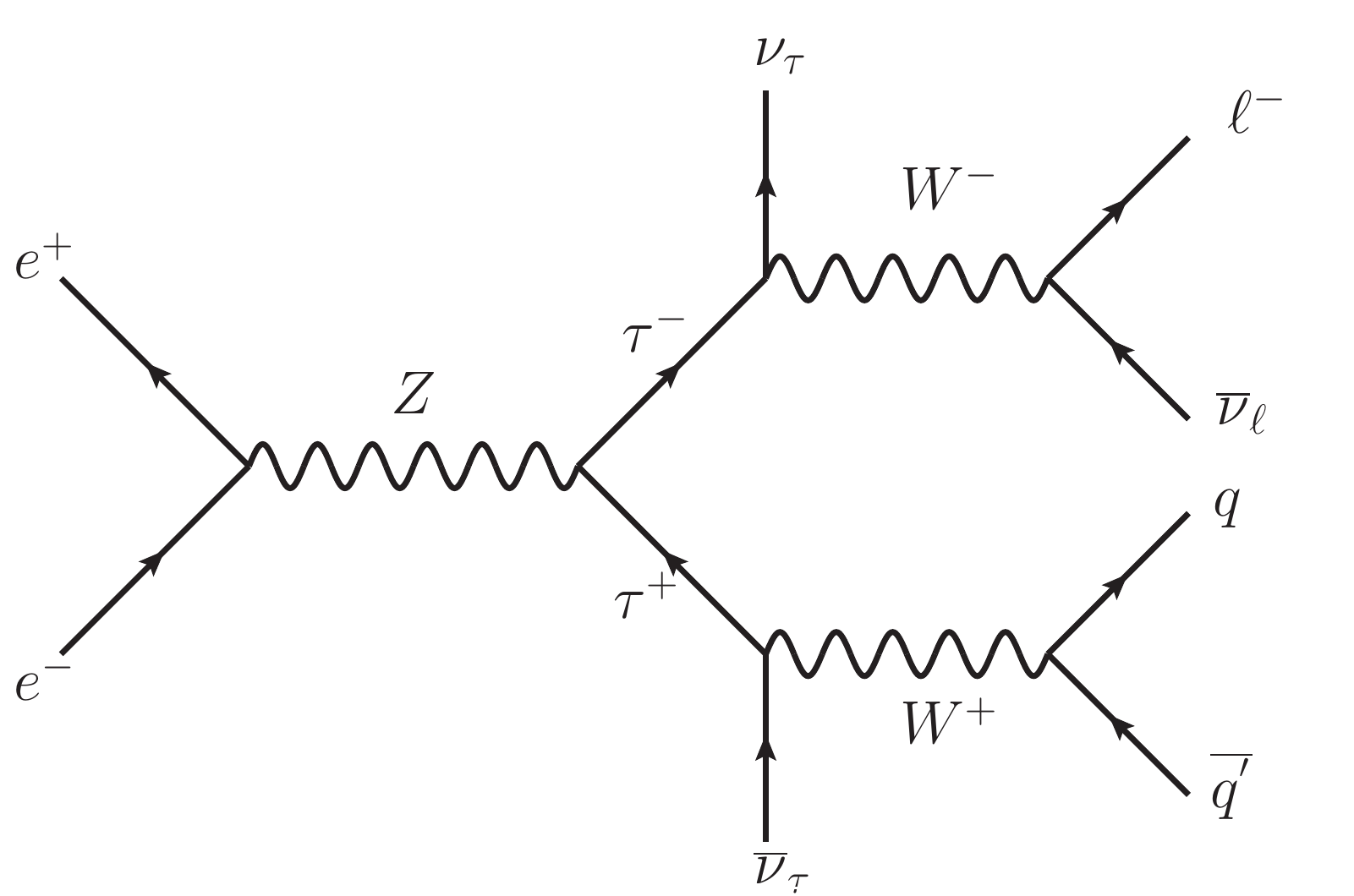}
\end{minipage}\begin{minipage}[c]{0.5\textwidth}
\vspace{-0.3cm}
\includegraphics[width=0.9\textwidth]{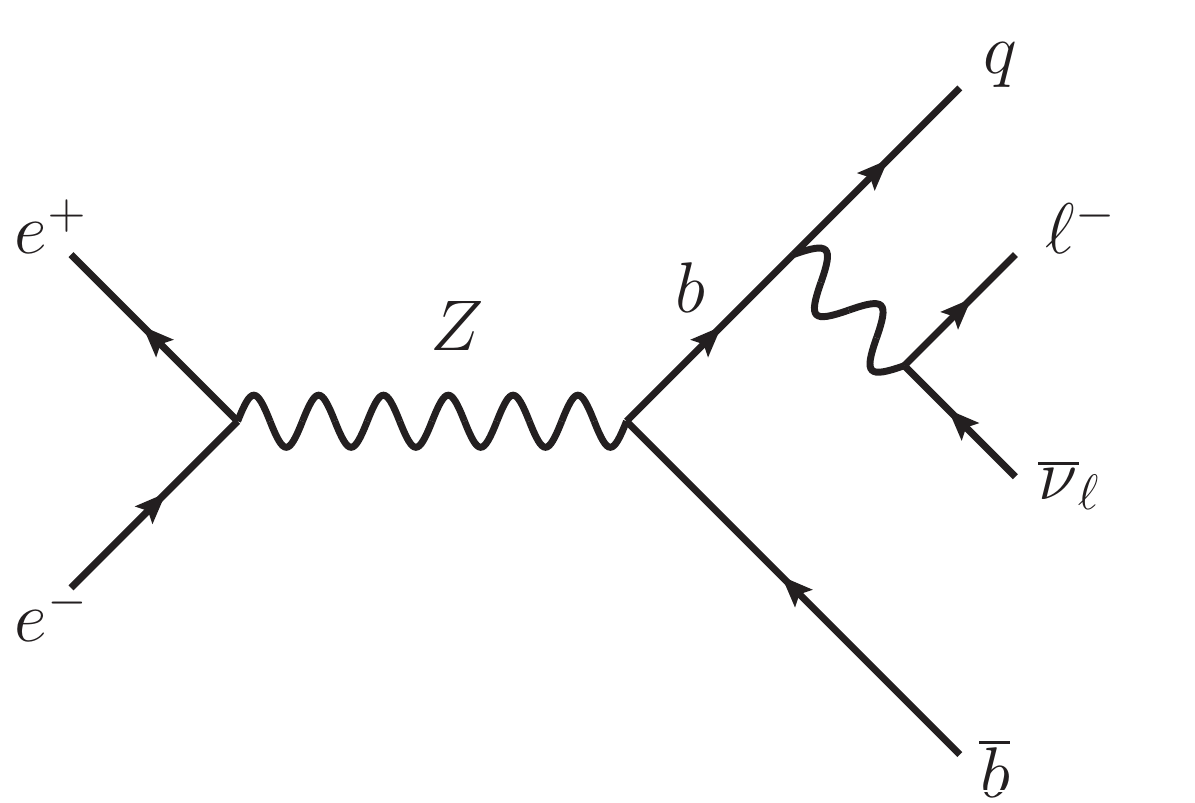}
\end{minipage}
\caption{Sample Feynman diagrams of the background.}\label{fig:bkg}
\end{figure}

The signal events are produced from the processes $e^+e^-\to \nu N\to \nu \ell jj$, and in this work we consider 
the charged lepton $\ell$ to be an electron or a muon. Therefore, the final states are forced into three jets 
(one leptonic jet and two hadronic jets for example) with the the $eekt$ algorithm. In both the electron and 
muon cases, the main background comes from the 
$e^{+}e^{-}\to jjjj$,  $b\overline{b}$ and $\tau^+\tau^-$ processes. In one $jjjj$ event, if one jet is too soft or collinear 
to the beam and is thus not detected but identified as missing energy, and simultaneously another jet is misidentified 
as an electron or muon, such an event may mimic a signal event. In this work, we assume that the misidentification rates of jets as 
electrons and muons are about $10^{-4}$. As for the $b\overline{b}$ and $\tau^+\tau^-$ events, the fermion pairs 
further decay into final states containing one charged lepton, two jets and missing energy, $\slashed{E}\ell jj$, 
as shown in FIG. \ref{fig:bkg}. It should be pointed out that, although the decay products of a $\tau$ lepton are usually 
recognized as one single jet, in the $eekt$-$exclusvie$ algorithm it is possible that the products are reconstructed 
into two jets. This is due to that this algorithm automatically cluster nearest objects step by step until the final state of 
the event is reconstructed to exactly 3 jets. As for the other $q\overline{q}$ events with $q$ being a lighter quark, production at 
the $Z$-mass pole makes them highly boosted such that it is more difficult for the decay products of $q$ or $\bar{q}$
to form two open jets. Therefore, it is much more difficult for a lighter quark pair $q\bar{q}$ 
to mimic a signal event than a $b\bar{b}$ pair. 

In the following, we will discuss how we choose the event selection conditions to suppress the background 
and increase the signal significance.
The $jjjj$ events that mimic signal events typically have missing energy with small transverse momenta $\slashed{E}_T$. 
Therefore, a cut requiring a least $\slashed{E}_T$ can efficiently suppress such background.

For the $\tau\tau$ background as shown in the left panel of FIG. \ref{fig:bkg}, the two $\tau$ leptons in an event fly back-to-back with a 
large boost. Therefore, the charged lepton and the total missing energy in the final state are almost collinear or 
reverse to each other, and the angle between the two jets decaying from the same $\tau$ are very small. 
These inspire some effective cuts on the angular distances $\Delta R_{jj}$ and $\Delta R_{\slashed{E}\ell}$, 
where $\Delta R = \sqrt{\Delta\eta^2+\Delta\phi^2}$ with $\Delta\eta$ and $\Delta\phi$ being the differences between  
the pseudorapidities and the azimuthal angles of the two involved objects, respectively.

In a $b\overline{b}$ background event as shown in the right panel of  FIG. \ref{fig:bkg}, 
the neutrino, the charged lepton and one of the two jets decay from the same bottom quark with 
a large boost, so not only the angular distance between the charged lepton and the neutrino
but also the angular distance between the jet and the neutrino are small. Therefore, we set cuts on $\Delta R_{\slashed{E}\ell}$ 
and $\Delta R_{\slashed{E}j}$ to reduce the background from the $b\overline{b}$ process.

Considering the signal process $e^+e^-\to \nu N$, the energy of the resolved light neutrino is fixed owing 
to the momentum-energy conservation, as 
\begin{align}
E_{\nu} = \frac{M_{Z}^{2}-M_{N}^{2}}{2M_{Z}}, 
\end{align}
which can also provide an effective cut to suppress the background. In practice, the missing energy in 
any event is reconstructed by summing the 3-momenta of the visible objects, $\slashed{E} = \left| \Sigma_i{\bf p}_i\right|$. 
If an event only contains one neutrino, $\slashed{E}$ is honestly the neutrino energy $E_\nu$ and 
the reconstructed total energy $E_\text{rec} = \slashed{E} + \Sigma_iE_i$ reproduces the collision energy, 
about 91.2 GeV. Otherwise, if an event contains more than one undetected object (only one 
undetected massive object also has such a property), the reconstructed total energy $E_\text{rec}$ may deviate from the collision energy. 
Therefore, we also use $E_\text{rec}$ cuts to suppress the corresponding background. 

Inspired by the above considerations and some practical tests, we choose the following event selection conditions 
for the three categories depending on different heavy neutrino mass, the small-mass range ($M_{N} < 65$ GeV), 
the middle-mass range (65 GeV $< M_{N} < 80$ GeV) and the large-mass range (80 GeV $< M_{N} < 85$ GeV).

\begin{itemize}
\item The event selection conditions for the small-mass range ($M_{N} < 65$ GeV): 
\begin{itemize}
\item $P_{T}^{j} > 5$~GeV, $|\eta_{j}| < 2$, $\Delta R_{jj} > 0.1$, btag $<$ 0.8, TauTag = 0, BTag = 0;
\item $P_{T}^{\ell} > 3$~GeV, $|\eta_{\ell}| < 1$;
\item $\slashed{E}_{T} > 20$ GeV, $|E_\text{rec} - M_Z| < 10$ GeV, $|M_{\ell jj} - M_N| < \Gamma^M_{1/2}$;
\item $1.0 < \Delta R_{\slashed{E}j} < 5.5$, $1.5 < \Delta R_{\slashed{E}\ell} < 5.0$;
\end{itemize}

\item the event selection conditions for the middle-mass range ($65 < M_{N} < 80$ GeV):
\begin{itemize}
\item $P_{T}^{j} > 5$~GeV, $|\eta_{j}| < 2$, $\Delta R_{jj} > 0.4$, btag $<$ 0.8, TauTag = 0, BTag = 0;
\item $P_{T}^{\ell} > 3$~GeV, $|\eta_{\ell}| < 1$;
\item $\slashed{E}_T > 10$ GeV, $|E_\text{rec} - M_Z| < 10$ GeV, $|\slashed{E}-\slashed{E}_0| < \Gamma^{\slashed{E}}_{1/2}$, 
$|M_{\ell jj} - M_N| < \Gamma^M_{1/2}$;
\item $1.0 < \Delta R_{\slashed{E}j} < 5.5$, $1.5 < \Delta R_{\slashed{E}\ell} < 5.0$;
\end{itemize}

\item the event selection conditions for the large-mass range ($80 < M_{N} < 85$ GeV):
\begin{itemize}
\item $P_{T}^{j} > 10$~GeV, $|\eta_{j}| < 2$, $\Delta R_{jj} > 0.4$, $M_{jj} > 55$~GeV, btag $<$ 0.8, TauTag = 0, BTag = 0;
\item $P_{T}^{\ell} > 3$~GeV, $|\eta_{\ell}| < 1$;
\item $\slashed{E}_T > 5$ GeV, $|E_\text{rec} - M_Z| < 10$ GeV, $|\slashed{E}-\slashed{E}_0| < \Gamma^{\slashed{E}}_{1/2}$, 
$|M_{\ell jj} - M_N| < \Gamma^M_{1/2}$;
\item $1.5 < \Delta R_{\slashed{E}j} < 5.5$, $1.5 < \Delta R_{\slashed{E}\ell} < 5.0$.
\end{itemize}
\end{itemize}

In general, we set cuts on the transverse momenta $P_{T}$ and the pseudorapidities $|\eta|$ 
of the charged lepton and the jets, the transverse missing energy $\slashed{E}_{T}$, the invariant 
mass of the two jets $M_{jj}$, the reconstructed total energy $E_\text{rec}$, the angular distances between 
the two jets $\Delta R_{jj}$, between each 
jet and the missing energy $\Delta R_{\slashed{E}j}$ and between the missing energy and the charged lepton 
$\Delta R_{\slashed{E}\ell}$. Note that the $P^j_T$ cut can not be chosen to be too low, because 
the Monte Carlo simulation does not fully model physics at very low scales. It should be totally safe 
if we set a cut of $P^j_T>$ 10 GeV, which we accept for the large-$M_N$ range. We have also attempted 
such a $P^j_T$ cut for the other two mass ranges, but find that in such a case the signal reconstruction efficiency 
is extremely suppressed. Therefore, the result with a $P^j_T>$ 10 GeV cut in the small- and middle-$M_N$ 
ranges would not honestly reflect the real sensitivity of a $Z$-factory, and we choose $P^j_T>$ 
5 GeV instead. In addition, from the signal simulation we get the central value of the detected missing 
energies $\slashed{E}_0$ and the half-height width $\Gamma^{\slashed{E}}_{1/2}$ of the distribution, and we 
require that the missing energy $\slashed{E}$ in each event does not lie beyond the region 
$\slashed{E}_0\pm\Gamma^{\slashed{E}}_{1/2}$. A sizable width $\Gamma^{\slashed{E}}_{1/2}$ is formed mainly 
because the energy resolution for jets and leptons is not perfect. We also require that the invariant mass of 
the lepton and the di-jet $M_{\ell jj}$ should lie in the region $M_{N}\pm\Gamma_{1/2}^{M}$, where $\Gamma_{1/2}^{M}$
is the half-width of the $M_N$ spectrum of the reconstructed signal events. To further suppress the background, we also use the btag, 
BTag and TauTag, which gives information of the probability of a jet being a $b$-jet, whether or not a jet has been
tagged as containing a heavy quark and whether or not a jet has been tagged as a tau, respectively. 

After the event selection, we find that the $jjjj$ events dominate the background for all the three $M_N$ 
ranges. Besides, while the $b\bar{b}$ and $\tau\tau$ contributions to the background are considerable, 
the other contributions like $c\bar{c}$, $jj\ell\ell$ and $\ell\ell\ell\ell$ are negligible.

%For the small-mass range, we find that the most efficient cut is the $\slashed{E}_{T}$ cut, which removes most of the background without touching most of signal events. In this case, the $\tau\tau$ and $b\bar{b}$ processes have similar contributions to the background. For the middle- and large-mass ranges, the most efficient cut to suppress the $\tau\tau$ background is the $M_j$ cut, and the most efficient one to suppress the $b\bar{b}$ background is the $\Delta R_{\slashed{E}j}$ cut. In these two cases, the $b\bar{b}$ processes dominate the background. 

\section{{\label{sec:level4}Results and analysis}}

In this section, based on the simulation of the signal and background events, we present the capacity of 
$Z$-factories to search for heavy neutrinos with three benchmark integrated luminosities $\mathcal{L}$, 
0.1 ab$^{-1}$, 1 ab$^{-1}$ and 10 ab$^{-1}$. In practice, we estimate the expected upper bounds on the 
cross sections $\sigma({e^{+}e^{-}\to\nu N\to \nu \ell jj})$ at 95\% confidence level (CL), which can be 
approximately obtained by solving the equation
\begin{align}\label{eq:sig}
\ s=\frac{N_{s}}{\sqrt{N_{B}+N_{s}}}=\frac{N_{s0}\times({\sigma}/{\sigma_0})}{ \sqrt{N_{B0}+N_{s0}\times ({\sigma}/{\sigma_0}) } }\sqrt{\frac{\mathcal{L}}{\mathcal{L}_{0}}},
\end{align}
with $s\approx$ 1.7. This equation is understood as follows. For convenience, we only perform the 
simulation once for each heavy neutrino mass, 
with a reference setup of a specific luminosity $\mathcal{L}_0$ and a specific cross section of the signal $\sigma_0$, 
which brings $N_{B0}$ detected background events and $N_{s0}$ detected signal events. Then, based 
on the simulation, other cases are obtained by scaling the luminosity and the cross section, as expressed 
in the second equality of the formula \eqref{eq:sig}.

\begin{figure}[h]
\begin{minipage}[c]{0.5\textwidth}
\includegraphics[width=1.1\textwidth]{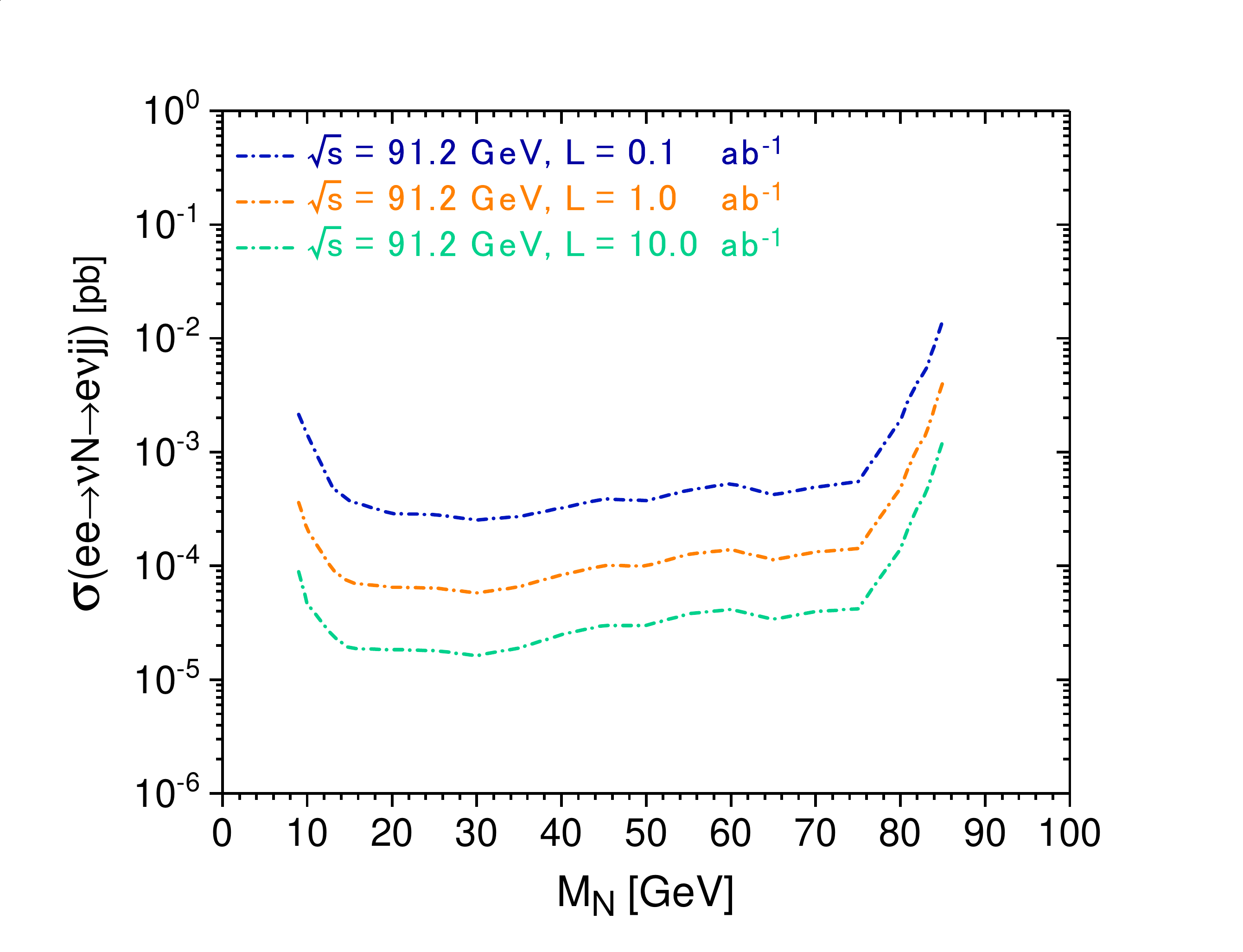}
\end{minipage}\begin{minipage}[c]{0.5\textwidth}
\includegraphics[width=1.1\textwidth]{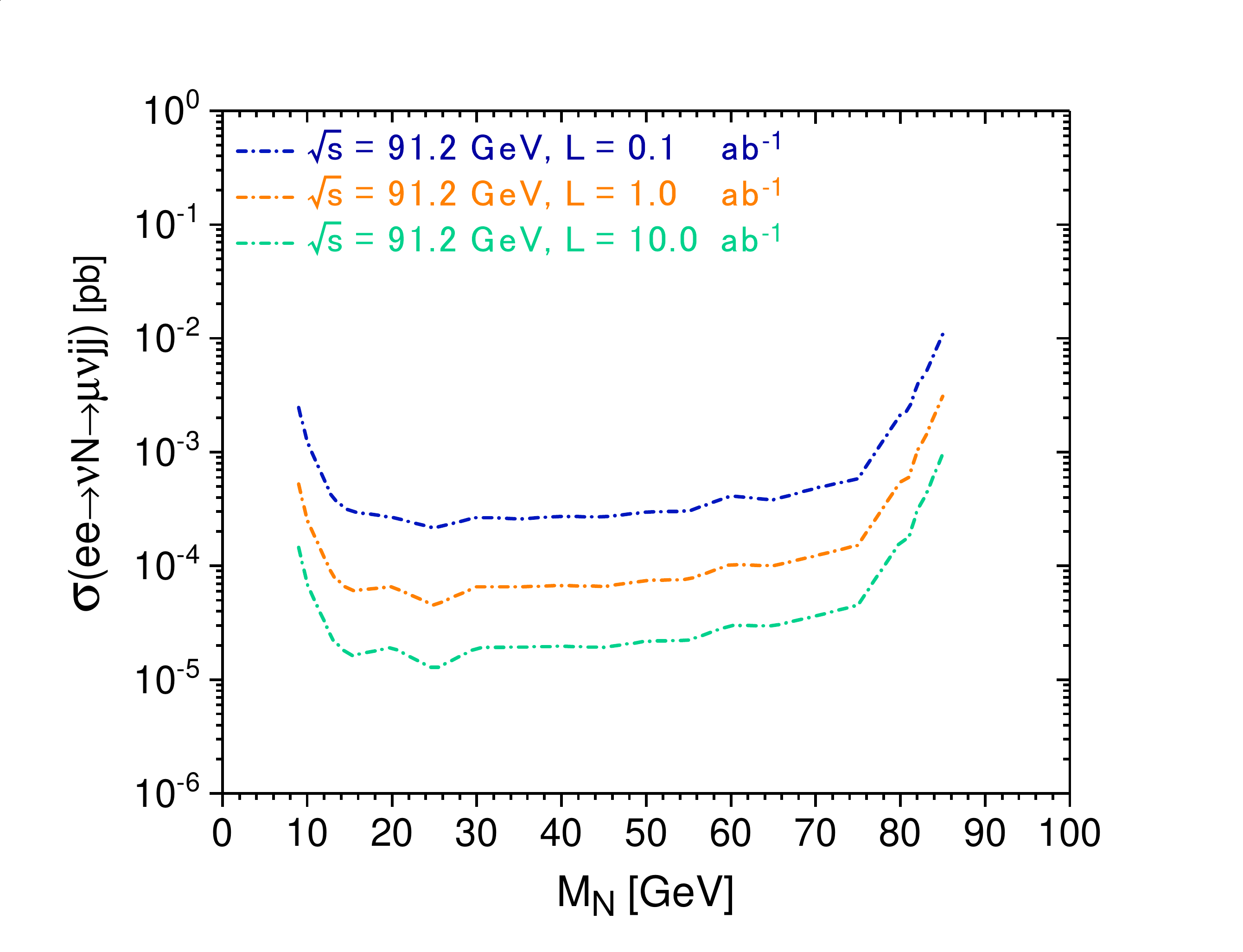}
\end{minipage}
\caption{The upper bounds on $\sigma({e^{+}e^{-}\to\nu N\to \nu e jj})$ (left) and 
$\sigma({e^{+}e^{-}\to\nu N\to \nu\mu jj})$ (right) by $Z$-factories at 95\% CL. 
The blue, orange and cyan curves correspond to integrated luminosities of 0.1 ab$^{-1}$, 
1.0 ab$^{-1}$ and 10 ab$^{-1}$, respectively. See text for details.} \label{fig:mainresult}
\end{figure}

\begin{figure}[h]
\begin{minipage}[c]{0.5\textwidth}
\includegraphics[width=1.1\textwidth]{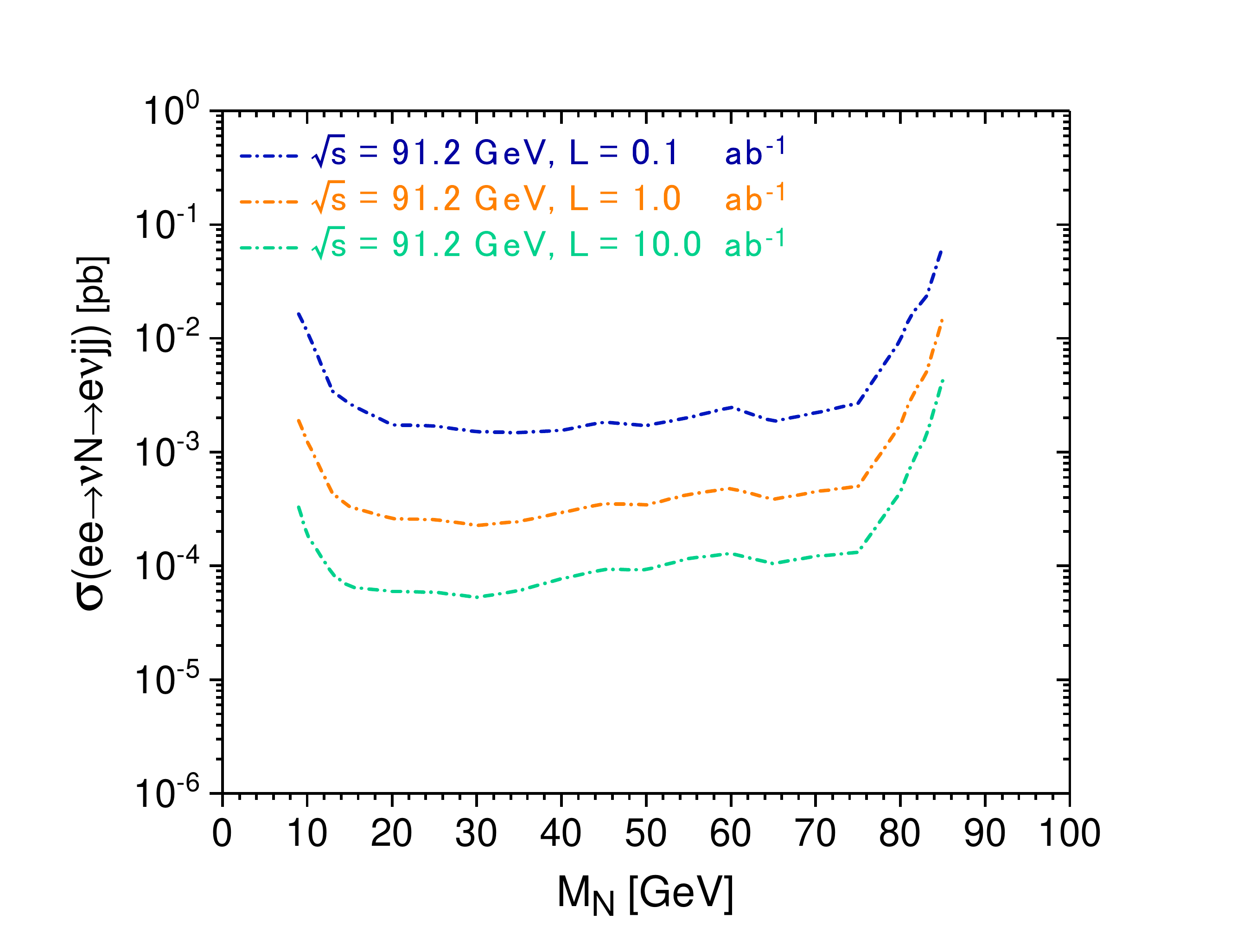}
\end{minipage}\begin{minipage}[c]{0.5\textwidth}
\includegraphics[width=1.1\textwidth]{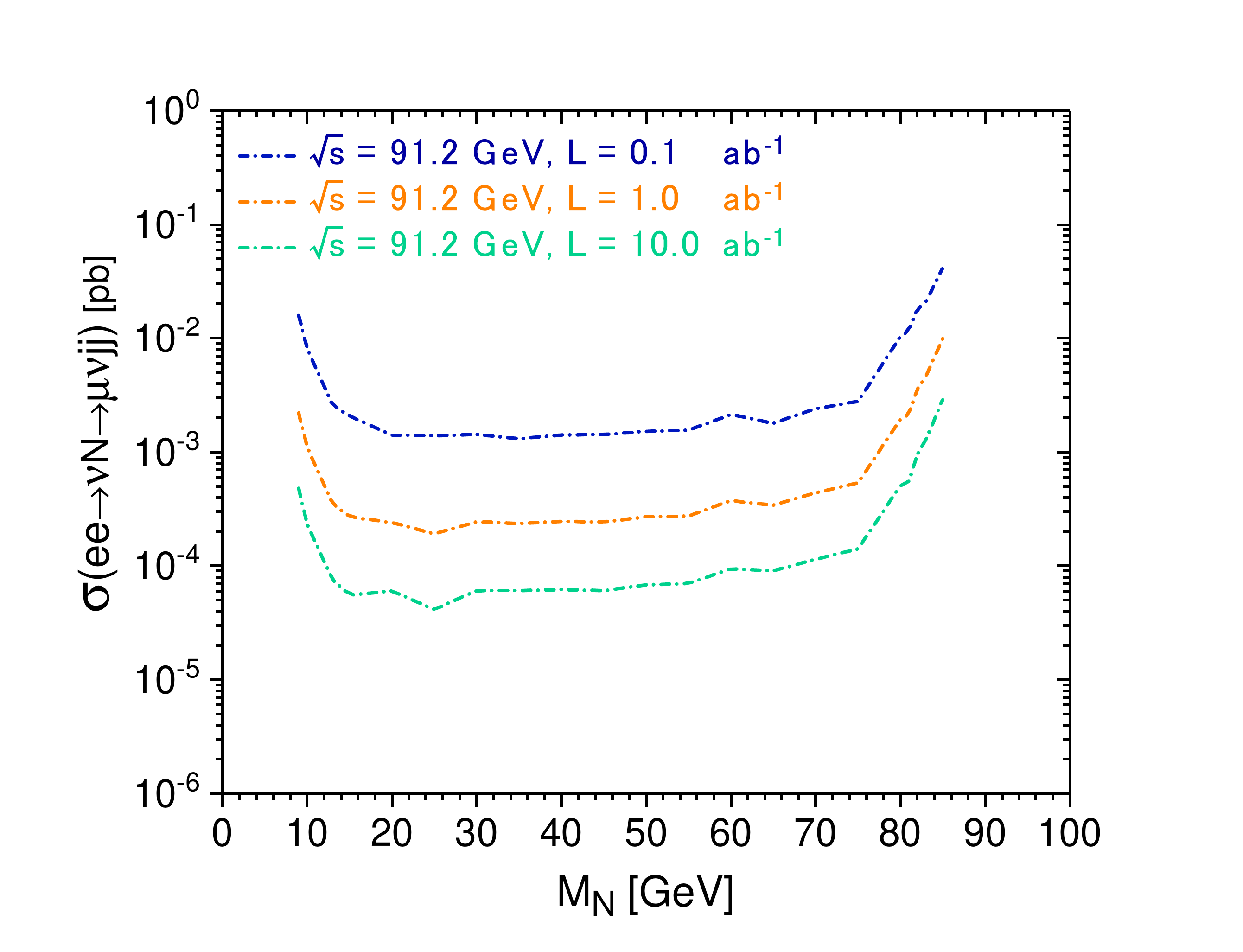}
\end{minipage}
\caption{The smallest $\sigma({e^{+}e^{-}\to\nu N\to \nu e jj})$ (left) and $\sigma({e^{+}e^{-}\to\nu N\to \nu\mu jj})$ 
(right) required to have heavy neutrinos discovered at $Z$-factories. The blue, orange and cyan curves correspond 
to integrated luminosities of 0.1 ab$^{-1}$, 1.0 ab$^{-1}$ and 10 ab$^{-1}$, respectively. See text for details.} 
\label{fig:mainresult-sig}
\end{figure}

The main results of this work, the upper bounds on $\sigma({e^{+}e^{-}\to\nu N\to \nu e jj})$ and 
$\sigma({e^{+}e^{-}\to\nu N\to \nu\mu jj})$ at 95\% CL given by future $Z$-factories, are shown in FIG. \ref{fig:mainresult}, with 
the heavy neutrino mass varying from 10 to 85 GeV. The curves for the integrated luminosities of 0.1 ab$^{-1}$, 
1 ab$^{-1}$ and 10 ab$^{-1}$ are presented. We find that for most of the mass range, {\it i.e.} 15 GeV $<M_N<$ 75 GeV, 
the upper bounds on the production cross sections are around a few $10^{-4}$ pb to $10^{-5}$ pb in both the electron 
and muon cases, with the integrated luminosities varying from 0.1 ab$^{-1}$, 1 ab$^{-1}$ and 10 ab$^{-1}$. 
Also, we assume that a significance $s$ larger than 5 in 
\eqref{eq:sig} indicates discovery of heavy neutrinos, and show the corresponding smallest discovery 
cross sections $\sigma({e^{+}e^{-}\to\nu N\to \nu e jj})$ and $\sigma({e^{+}e^{-}\to\nu N\to \nu\mu jj})$ 
in FIG. \ref{fig:mainresult-sig}.

In principle, the measurement of the total width of the $Z$ boson $\Gamma_Z=2.4952\pm0.0023$ GeV 
\cite{Tanabashi:2018oca} also sets constraints to the $e^{+}e^{-}\to Z\to\nu N$ cross sections, but we find 
that heavy neutrinos with their signal cross sections below $Z$-factory bounds in FIG. \ref{fig:mainresult} 
automatically satisfy the $\Gamma_Z$ constraints. For example, we consider a heavy neutrino with 
$\sigma({e^{+}e^{-}\to\nu N\to \nu\ell jj})$ of $\mathcal{O}(10^{-2})$ pb, which is the largest allowed cross 
section if no signals are observed at a $\mathcal{L}$ = 0.1 ab$^{-1}$ machine. It corresponds to an additional 
contribution $\Gamma(Z\to\nu N\to \nu\ell jj)$ of $\mathcal{O}(10^{-7})$ GeV to the $Z$ decay width. If 
the branching ratio $\mathcal{B}(N\to\ell jj)$ is of $\mathcal{O}$(10\%), then the existence of such a 
heavy neutrino will enlarge the total $Z$ decay width by $\Gamma(Z\to\nu N)$ of $\mathcal{O}(10^{-6})$ 
GeV, which is smaller than the uncertainty of the data $\Gamma_Z=2.4952\pm0.0023$ GeV by three orders. 
In other words, the constraint from the measurement of $\Gamma_Z$ is not comparable with the constraints 
given by this work.

\begin{figure}[h]
\begin{minipage}[c]{0.5\textwidth}
\includegraphics[width=1.1\textwidth]{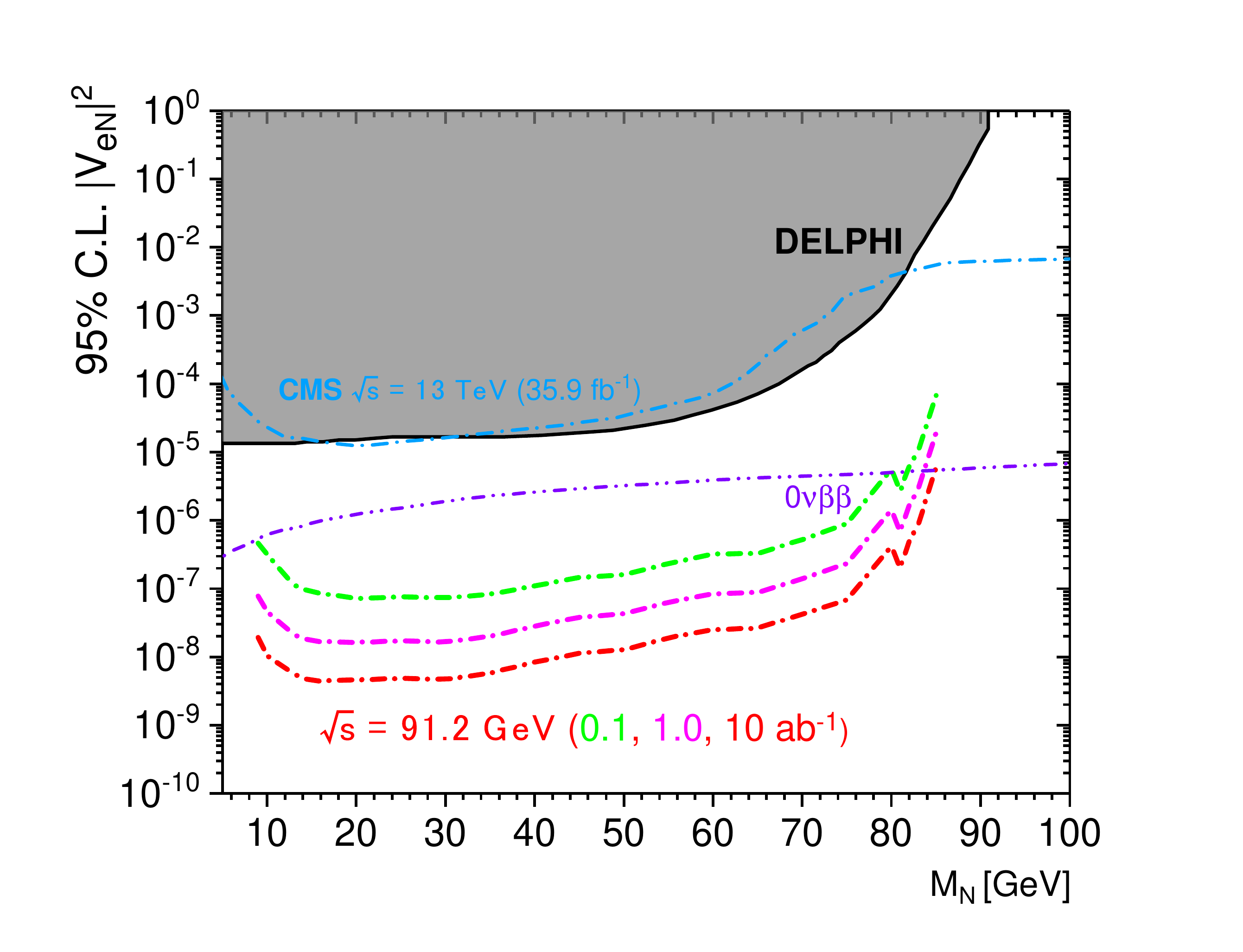}
\end{minipage}\begin{minipage}[c]{0.5\textwidth}
\includegraphics[width=1.1\textwidth]{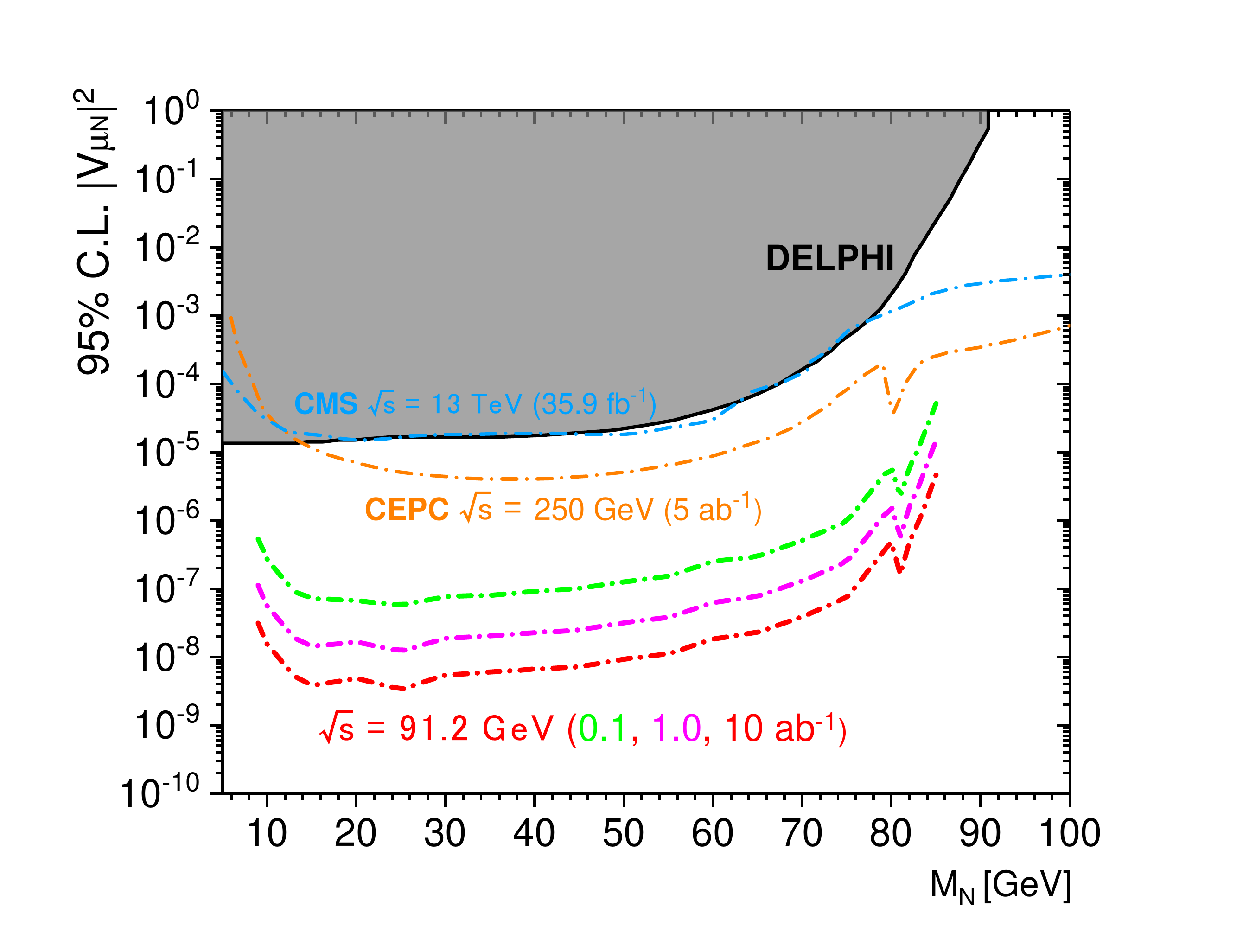}
\end{minipage}
\caption{The upper bounds on the mixing parameters $|V_{eN}|^{2}$ (left) and $|V_{\mu N}|^{2}$ (right) 
given by $Z$-factories at 95\% CL, compared to the upper bounds given by DELPHI \cite{Abreu:1996pa} and 
CMS \cite{Sirunyan:2018mtv}, 
the $0\nu2\beta$ decay experiments \cite{Elliott:2004hr,Benes:2005hn,Rodejohann:2011mu} and the 
CEPC as a Higgs factory \cite{Zhang:2018rtr}. The green, pink and red curves correspond to integrated 
luminosities of 0.1 ab$^{-1}$, 1.0 ab$^{-1}$ and 10 ab$^{-1}$, respectively. See text for details.}\label{fig:mixpara}
\end{figure}

To have a direct comparison with previous experimental constraints and some relevant future ones, we consider 
the case in which \eqref{eq:lagrangian} is the only source of new dynamics beyond the SM and give the $Z$-factory 
constraints on the mixing parameters $|V_{eN}|^2$ and $|V_{\mu N}|^2$ in FIG. \ref{fig:mixpara}.\footnote{We are 
aware that in such a case $Z$-factories are able to give much more stringent constraints on these mixing parameters 
than our constraints by making use of displaced vertex information \cite{Antusch:2016vyf,Antusch:2017pkq}.} This is 
achievable because the cross sections are proportional to the corresponding mixing parameters, 
$\sigma({e^{+}e^{-}\to\nu N\to \nu \ell jj})\propto|V_{\ell N}|^{2}$, as analysed in the last paragraph of Section 
\ref{sec:level2}. For both $|V_{eN}|^2$ and $|V_{\mu N}|^2$, we find a large improvement compared to DELPHI 
\cite{Abreu:1996pa} and CMS \cite{Sirunyan:2018mtv}, the upper bounds being decreased typically by two orders
of magnitude even with the lowest luminosity 
setup. For $|V_{eN}|^2$, the given upper bounds by the considered $Z$-factories are lower than that given by the 
$0\nu2\beta$ decay experiments \cite{Elliott:2004hr,Benes:2005hn,Rodejohann:2011mu} by at least one order of magnitude 
in most of the mass range of the heavy neutrino. While for $|V_{\mu N}|^2$, the upper bounds given by the $Z$-factories 
are at least two orders of magnitude lower than that given by the CEPC as a Higgs factory \cite{Zhang:2018rtr} when 
$M_N<$ 70 GeV.
One may worry that a heavy neutrino with mixing parameters as small as such bounds will be so stable that the 
detection of its decay is always out of reach by detectors. If this is true, the bounds given in FIG. \ref{fig:mixpara} 
will not be valid, since all our analyses are based on the assumption that the signal events can be detected within 
detectors once they happen. To clarify that this will not be a problem, we estimate how far a 10 GeV heavy neutrino 
can typically fly before its decay, and the flying distances of the other heavier neutrinos are always shorter given the 
same relevant mixing parameters. Considering a detector having a diameter of $\mathcal{O}$(1) meter, we find that 
as long as the mixing parameter $\sum_{\ell}|V_{\ell N}|^{2}$ is larger than $\mathcal{O}(10^{-9})$, the 10 GeV heavy 
neutrino is most likely to decay before it can fly out of the detector. Comparing $\mathcal{O}(10^{-9})$ with the 
bounds in FIG. \ref{fig:mixpara}, one can find that the bounds in the cases with $\mathcal{L}$ = 0.1 and 1 ab$^{-1}$ 
are not affected by the limited size of the detector, and that the case with $\mathcal{L}$ = 10 ab$^{-1}$ is also basically safe.

\section{\label{sec:level5}Conclusion}

To conclude, we have presented a study of possible heavy neutrino searches at future $Z$-factories. For different 
heavy neutrinos with mass ranging from 10 to 85 GeV, we have obtained the expected upper bounds on the 
production cross sections of their discovery processes $e^{+}e^{-}\to\nu N\to \nu e jj$ and $e^{+}e^{-}\to\nu N\to \nu\mu jj$ 
given by $Z$-factory with $\mathcal{L}$ = 0.1, 1 and 10 ab$^{-1}$, respectively. Under the assumption that 
the interactions between the heavy neutrinos and the SM particles are only induced by the neutrino mixing, 
the constraints on the cross sections have been translated to the constraints on the corresponding mixing 
parameters $|V_{eN}|^2$ and $|V_{\mu N}|^2$, which, depending on the luminosity setup, are typically improved 
by two to four orders compared to the DELPHI constraints \cite{Abreu:1996pa} and the CMS constraints 
\cite{Sirunyan:2018mtv}. We also find that the future 
$Z$-factories will set much more stringent constraints on $|V_{eN}|^2$ than the $0\nu2\beta$ decay experiments 
\cite{Elliott:2004hr,Benes:2005hn,Rodejohann:2011mu} by one to three orders, and on $|V_{\mu N}|^2$ than the 
CEPC as a Higgs factory \cite{Zhang:2018rtr} by two to three orders, given the heavy neutrino mass is smaller than 80 GeV.

\section*{Acknowledgements}

This work is supported by the Fundamental Research Funds for the Central Universities under the Grant 
No.~lzujbky-2019-it08, the National Natural Science Foundation of China under the Grant No.~U1732101, 
the Gansu Natural Science Fund under the Grant No.~18JR3RA265 and
the Deutsche Forschungsgemeinschaft (DFG) within research unit FOR 1873 (QFET). We are grateful to Qing-Hong Cao, Cheng Chen, Min He, Qiang Li, Qi-Shu Yan and Ye-Ling Zhou for 
useful discussions. In addition, JND would like to thank Ying Guan for her encouragement.

\end{document}